\documentclass[prl,twocolumn,showpacs,nofootinbib]{revtex4}
\usepackage{graphicx}
\usepackage{epstopdf}
\usepackage{epsfig}
\usepackage{color}

\newcommand{\be}{\begin{equation}}
\newcommand{\ee}{\end{equation}}
\renewcommand{\r}{{\bf r}}

\begin{document}

%----------------------------------------------------------------------------

\title{Cooperative effects  in one-dimensional random atomic gases: \\ Absence of  single atom limit}
\author{E. Akkermans$^{1}$ and A. Gero$^{1,2}$}
\affiliation{$^{1}$Department of Physics, Technion-Israel Institute of Technology,
  32000 Haifa, Israel \\ $^{2}$Department of Education in Technology and Science, Technion-Israel Institute of Technology, 32000 Haifa, Israel}
%--------------------------------------------------------------------------

\begin{abstract}
We study superradiance in a one-dimensional geometry,  where $N\gg 1$ atoms are randomly distributed along a line. We present an analytic calculation of the photon escape rates based on the diagonalization of the $N \times  N$ coupling matrix $U_{ij}=\cos x_{ij}$, where $x_{ij}$ is the dimensionless random distance between any two atoms.  We  show that unlike a three-dimensional geometry, for a one-dimensional atomic gas the single-atom limit is never reached and the photon is always localized within the atomic ensemble. This localization originates from long-range cooperative effects and not from disorder as expected on the basis of the theory of Anderson localization.

\end{abstract}


\maketitle
%-----------------------------------------------------------------------

\section{I. Introduction}
Cooperative effects such as superradiance and subradiance  \cite{dicke, gross} originate from  indirect interactions between atoms through the radiation field. These effects show up as a multi-atomic coherent emission which is qualitatively  different from that of a single atom \cite{stephen,lehmberg}. Cooperative effects have  been studied both theoretically and experimentally in various systems, such as quantum dots \cite{scheibner}, Bose-Einstein condensate \cite{inouye,baumann}, cold atoms \cite{bienaime,kuraptsev} and Rydberg gases \cite{wang}.

 In the context of cold atoms, photon localization, which occurs as a decrease of photon escape rates from disordered media, has been investigated \cite{PRL2}. It has been shown that for a three-dimensional atomic system, photon localization is primarily determined by cooperative effects rather than by disorder. Moreover, localization shows up as a  crossover between delocalized and localized photons and not as a disorder-driven phase transition as for Anderson localization. 

In this Letter we study photon escape rates from a disordered one-dimensional atomic gas and compare them to those obtained in a three-dimensional geometry.  We will  show that unlike a three-dimensional geometry, for a one-dimensional atomic gas the single atom limit is never reached and the photons are always localized. This localization stems only from long-range cooperative effects and not from disorder as expected from the theory of Anderson localization.

\section{II. Model}
We are interested in the dipolar interaction of $N\gg 1$ identical atoms with a scalar radiation field.
%\be E(\textbf{r})=i\sum_{\bf{k}}\sqrt{\frac{\hbar\omega_{k}}{2\epsilon_{0}\Omega}}(a_{\bf{k}}
%e^{i\bf{k}\cdot r}-a_{\bf{k}}^{\dag}e^{-i\bf{k}\cdot r})\label{Sc}.\ee
%$a_{\bf{k}\varepsilon}$ and $a_{\bf{k}\varepsilon}^{\dag}$ are, respectively,
%the annihilation and creation operators of a mode of the field
%of wave vector $\bf{k}$ and angular frequency $\omega_{k}=ck$.
%$\Omega\ $ is a quantization volume, $\epsilon_{0}$ is the vacuum dielectric constant, and $c$ is the speed of light in %vacuum.
Here, atoms are taken as non-degenerate, two-level systems. The energy
separation between the excited state $|e \rangle$ and the ground state $|g
\rangle$, including radiative shift,  is
$\hbar\omega_{0}$ and the inverse lifetime of the excited level is
$ \Gamma_{0}$. Indeed, this two-level atom model neglects
the the energy structure of a real atom, but as
selection rules restrict the allowed transitions between
states, this approximation is more than a mathematical convenience.

We consider a one-dimensional geometry where the atoms are randomly distributed along a line.  Moreover, only modes of field that belong to an elongated pencil-shaped radiation pattern parallel to the inter-atomic axis are taken into account. This radiation pattern, obtained in a pencil-shaped cavity, corresponds to the directional emission along the cavity axis \cite{tallet1,milonni2}.

We neglect  recoil effects  and the Doppler shift by assuming that the typical speed of the atoms  is large compared to $\hbar k / \mu$  but small compared to $\Gamma_{0}/k$ where $k$ is the radiation wavenumber and $\mu$ is the mass of the atom.  Additionally, we neglect retardation effects so each atom is allowed to influence the others instantaneously.

\section{III. Dicke states and Cooperative effects}
The absorption of a photon by a pair of atoms, each in its ground state and, respectively, located at $\r_{1}$ and $\r_{2}$, leads to a configuration where one atom is excited  while the other is de-excited. The possible  configurations can be represented by the Dicke states \cite{dicke}. The singlet Dicke state is \be |-\rangle=|00
\rangle = {1 \over \sqrt{2}} [ |e_{1} g_2 \rangle - |g_{1} e_{2}
\rangle ] \label{eq8a}\ee and the triplet Dicke states are
%\be
\begin{eqnarray}
%{c}
|11 \rangle &=& |e_{1} e_2  \rangle,  \nonumber \\ |+ \rangle = |10 \rangle &=& {1 \over \sqrt{2}} [ |e_{1} g_2 \rangle + |g_{1} e_{2} \rangle], \nonumber  \\ |1-1\rangle &=&|g_{1} g_2 \rangle.
\label{eq8b}
\end{eqnarray}
%\ee
%The states $|11 \rangle$ and $|1-1 \rangle$ correspond,
%respectively, to both atoms in their excited states and both atoms
%in their ground state. The singlet state $|-\rangle=|00 \rangle$ and the
%triplet state $|+\rangle=|10 \rangle$ both correspond to one atom in the
%excited state and the other in the ground state, but $|- \rangle$
%is anti-symmetric  where $|+ \rangle$ is symmetric under an
%exchange of the atoms.

These states are characterized by an effective interaction potential and  a
modified lifetime as compared to independent atoms.
For the one-dimensional geometry considered here,  the cooperative spontaneous emission rate or the inverse lifetime  of the states $|\pm>$  is \cite{milonni2} \be {\Gamma^\pm \over \Gamma_{0}}
= 1\pm \cos k_{0}r \label{eqa37},\ee where  $k_{0}=\omega_{0}/c$ and $r=|\r_{1}-\r_{2}|$.
The corresponding cooperative radiative level shift or the interaction potential  is given by
$\Delta E^{\pm} =  \pm \frac{\hbar}{2}  \Gamma_{0} \sin k_{0}r \label{eqa38}.$

For comparison, in a three-dimensional  system, the cooperative spontaneous emission rate is \cite{vries} \be {\Gamma^\pm \over \Gamma_{0}}= 1\pm \frac{\sin k_0r}{k_0 r},\label{SF2}\ee  and the  corresponding cooperative radiative level shift   is given by
$ \Delta E^{\pm} =  \mp  \frac{\hbar}{2} \Gamma_{0} \cos k_{0}r /  {k_{0}r} \label{SF3}.$ 

%For a one-dimensional geometry, the cooperative spontaneous emission rate, $\Gamma'^{\pm}$, is \cite{milonni2} \be $%{\Gamma'^\pm \over \Gamma_{0}'}
%= 1\pm \cos k_{0}r \label{eqa37},\ee where the one-dimensional spontaneous emission rate is $ $%\Gamma_{0}'=d^{2}\omega_{0}/\hbar\epsilon_{0}c$ and $d$ is the reduced matrix element of the electric dipole %moment $operator of an atom.

When the atoms are close enough $(k_{0}r\ll1)$, the Dicke limit is obtained in both geometries, namely,      
 $\Gamma^{\pm}=(1\pm1)\Gamma_{0}$. But, when the atoms are well separated  $(k_{0}r\gg1)$, the single-atom limit  is not recovered in eq.~(\ref{eqa37}) since the one-dimensional inverse lifetime is a periodic function of the inter-atomic distance, while the three-dimensional one  falls off with the inter-atomic separation. Similarly, the range of the one-dimensional interaction potential  is infinite, while it is finite in the three-dimensional case. This fundamental difference will be the driving effect in the calculation of photon escape rates from an atomic gas in the next section.

\section{IV. Photon escape rates from atomic gases}
To go beyond the case of two atoms, we follow \cite{PRL2, tallet1} who
studied the equation of motion for the reduced atomic density operator $\rho$ of  a gas of atoms with a single excitation.
The time evolution of the ground state population associated with $\rho$ is given by
\be \frac{d\langle G| \rho |G \rangle}{dt}= \Gamma_{0}\sum_{ij}U_{ij}\langle G|S_{j}^{-}\rho S_{i}^{+}|G\rangle \label{evo},\ee
where $|G\rangle=|g_{1}, g_{2},...,g_{N}\rangle$ and $S_{i}^{\pm}$ is the raising (lowering) operator of atom $i$.
$U$ is an $N \times N$ Euclidean random matrix as defined hereafter.
For the one-dimensional geometry \be U_{ij}=\cos k_0 r_{ij},\label {U}\ee while for the three-dimensional gas
 \be U_{ij}=\frac{\sin k_0r_{ij}}{k_0r_{ij}},\label {u}\ee  where $r_{ij}=|\r_{i}-\r_{j}|$ is the random distance between any two atoms. 
With the help of the eigenvalue equation of $U$, namely $\sum_{j=1}^{N}U_{ij}u_{j}^{(n)}=\Gamma_{n}u_{i}^{(n)}$, we  rewrite eq.~(\ref{evo}) as
$ d\langle G| \rho |G \rangle/dt= \Gamma_{0}\sum_{n=1}^{N}\Gamma_{n}\langle G|S_{n}^{-}\rho S_{n}^{+}|G\rangle$
where $S_{n}^{\pm}=\sum_{i=1}^{N}u_{i}^{(n)}S_{i}^{\pm}$ are the collective raising and lowering operators.
Thus, we can interpret the eigenvalues $\Gamma_{n}$  of the coupling matrix $U$ as the photon escape rates from the gas and define the dimensionless average density of photon escape rates  as
\be P(\Gamma)=\frac{1}{N}\overline{\sum_{n=1}^{N}\delta(\Gamma-\Gamma_{n})}\label{eqc30},\ee where the average, denoted by $\overline{\cdot
\cdot \cdot}$, is taken over the spatial configurations of the atoms.
%These results are not surprising since eq.~(\ref{U}) and eq.~(\ref{u}) are just the Green's functions for a  scalar photon %in a one- and three-dimensional system, respectively.

The quantity  used as a measure of photon localization is  the  normalized function $C=1-2\int_1^\infty d\Gamma P(\Gamma)$. $C$ thus measures the relative number of states having a vanishing escape rate. In the three-dimensional case, discussed in \cite{PRL2}, the function $C$ exhibits a scaling behavior over a broad range of system size and disorder. The scaling variable is $N / N_\perp$, where $N_\perp$  is the number of transverse photon modes in the system. $C$ allows to compare the different contributions of disorder and cooperative effects to photon localization, although an unambiguous distinction between the two mechanisms can not be achieved in the three-dimensional geometry \cite{PRL2}.  
In this Letter, we will show that in the one-dimensional case $C$ exhibits a scaling behavior as well, and the scaling variable  is $N$. We  will also show that the expression of $C$ obtained in the one-dimensional geometry is valid for both ordered and disordered systems, thus  it is possible  to unambiguously discern between the  contributions of disorder and cooperative effects to photon localization.

The eigenvalues of $U$ in eq.~(\ref{u}) have been obtained numerically in \cite{PRL2}. %It has been shown that photon localization occurs as a smooth crossover between delocalized and localized photons.%
 Recently, based on the Marchenko-Pastur law \cite{MPL}, the authors of \cite{skipetrov} have approximated  the spectrum of  this  matrix  in the limit of large systems. Here, we will provide an  analytic diagonalization of (\ref{U}) that holds for an arbitrary system size. 
%We will also show that unlike for a three-dimensional geometry, in a one-dimensional atomic gas the single atom limit is %never reached and the photons are always localized. 

\begin{figure}[ht]
\centerline{ \epsfxsize 9cm \epsffile{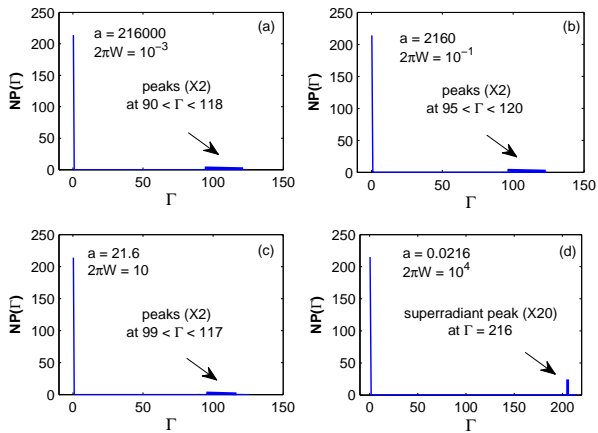} } \caption { Behavior of $NP (\Gamma)$ for different values of $W$, $a$ and for $N = 216$ in a one-dimensional geometry. The single atom limit is never reached and the photon is always localized in the atomic gas. Panels (a)-(c) describe  large samples. The  Dicke limit is shown in (d).}
\label{fig1}
\end{figure}

To that purpose, we  consider $N\gg1$ atoms confined in a one-dimensional system of length $L=2\pi a/k_0$. The atoms are randomly distributed with a uniform density $N/L$ and the corresponding coupling matrix is given by eq.~(\ref{U}).
The average density of photon escape rates, obtained for many random configurations of the atoms, is presented in fig. \ref{fig1} for different values of $W=N/2\pi a$ and $a$. A remarkable difference between the one and the three-dimensional case \cite{PRL2} is observed. Unlike the three-dimensional geometry, the single atom limit is never reached and the photons are always localized in the atomic gas.
 % However, Dicke limit is the same in both geometries.

%------------------------------
%------------------------------

Let us distinguish between two regimes, Dicke regime where $a\ll1$  and the large sample regime, where $a\geq1$. In Dicke regime the coupling matrix is \be
U = \left(\begin{array}{llllllll} 1 & 1 & \cdots & 1 \\
1 & 1 & \cdots & 1 \\
\vdots & \vdots &  & \vdots \\
1 & 1 & \cdots & 1
\end{array}\right) \
\label{matrice}
.\ee
Thus, the average density of photon escape rate is given by \be P(\Gamma)=\frac{1}{N}[(N-1)\delta(\Gamma)+\delta(\Gamma-N)]\label{eqc33},\ee  as presented in fig. 1(d). Eq.~(\ref{eqc33}) holds in  Dicke regime of the three-dimensional case as well.
The spectrum of $U$ given above yields $C=1-2/N$. For the current case where $N\gg 1$, $C=1$ indicating photon localization.

Away from the Dicke limit, in the large sample regime shown in fig. 1(a)-(c), $ P(\Gamma)$ is calculated as follows. The $N\times N$ matrix $ U_{ij}=\cos k_{0}r_{ij}$ may be rewritten as
$U=\frac{1}{2} A^{\dag}A$, where $A$ is the $2\times N$ matrix defined by
$ A_{0j}=e^{ik_0r_j}$ and
$A_{1j}=e^{-ik_0r_j}.$
As $U$ is a real symmetric matrix,  its non-vanishing eigenvalues can be found from those of $ U^\dag$,  given by
\be
U^\dag =\frac{1}{2} \left(\begin{array}{ll} N & M  \\
M^{*} & N  \\
\end{array}\right)\label{Udag}. \
\ee
Here $M=\sum_{k=1}^{N}e^{2ik_0r_k}\label {M}$ is a random variable where $k_0r_k$ is uniformly distributed over $[0,2\pi a]$. Since the two eigenvalues of eq.~(\ref{Udag}) are
\be \lambda_\pm=\frac{N\pm|M|}{2} \label{lambda},\ee
the spectrum of $U$ is given by \be P(\Gamma)=\frac{1}{N}\left[ (N-2)\delta(\Gamma)+\delta(\Gamma-\lambda_+)+\delta(\Gamma-\lambda_-)\right].\label{spec}\ee
We can estimate $|M|$ by writing
\be |M|^2=N+\sum_{p\neq q}e^{2ik_0(r_p-r_q)},\ee where the second term involves $N(N-1)$ terms. On average over non-correlated disorder the second term vanishes so that $|M|\sim \sqrt{N}$.
For the spectrum of $U$ given in eq.~(\ref{spec}) it is evident that $C=1-4/N$. Thus,
for large values of $N$ the photons are localized in the gas.

\begin{figure}[ht]
\centerline{ \epsfxsize 9cm \epsffile{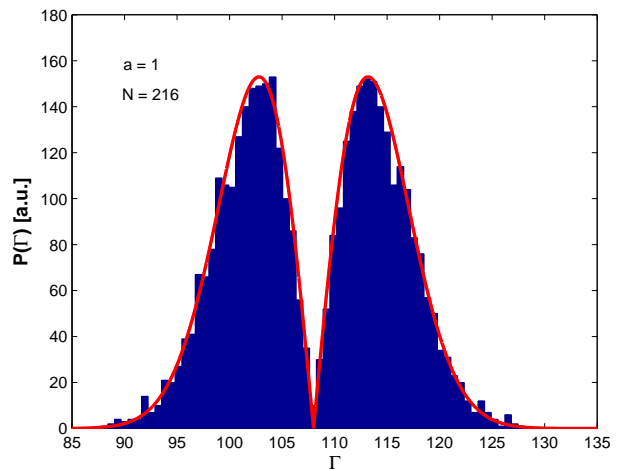} } \caption { Distribution of the eigenvalues  of $U_{ij}=\cos k_0 r_{ij}$, obtained numerically for $a=1$ $ (k_0L=2\pi)$ and $N=216$. The degenerate subradiant mode at $\Gamma=0$ is not presented.  The two superradiant modes are centered at $\Gamma= (N\pm \sqrt{N/2})/2$. The solid line is calculated  by eq.~(\ref{lambda}) and the Rayleigh distribution given  in eq.~(\ref{sc}).}
 \label{fig2}
\end{figure}

In order to calculate exactly the distribution function of $|M|$, first we assume that $a$ is an integer. In this special case, the distribution function is just the Rayleigh distribution,
\be P(|M|)=\frac{2|M|}{N} e^{-\frac{|M|^2}{N}},\label{sc}\ee whose mode is $\sqrt{N/2}$.
Figure \ref{fig2} shows the eigenvalue distribution spectrum  of $U$  for $a=1$ (excluding the degenerate subradiant mode at $\Gamma=0$) as well as the calculated $P(\Gamma)$ given by eqs.~(\ref{lambda}) and (\ref{sc}).

%------------------------------
%------------------------------

%------------------------------
\begin{figure}[ht]
\centerline{ \epsfxsize 9cm \epsffile{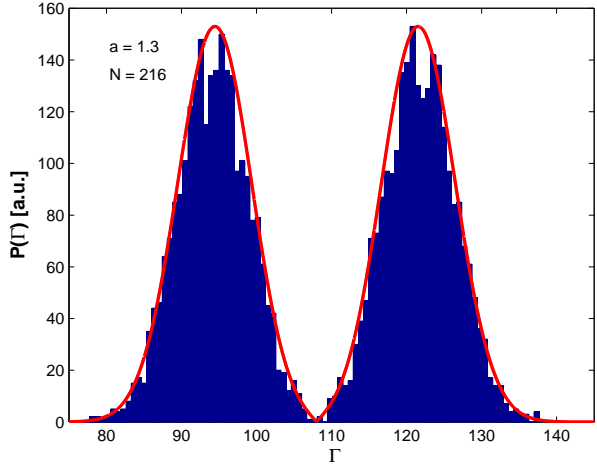} } \caption { Distribution of the eigenvalues  of $U_{ij}=\cos k_0 r_{ij}$, obtained numerically  for $a=1.3$ and $N=216$. The degenerate subradiant mode at $\Gamma=0$ is not presented. The solid line is calculated  by eqs.~(\ref{lambda})  and (\ref{beck}).}
 \label{fig3}
\end{figure}
%------------------------------

In the general case, for an arbitrary value of $a$, we follow \cite{beckmann}, as described below. As $N\gg1$, according to the Central Limit theorem, $\mbox{Re(\emph{M})}$ is normally distributed with mean $m_1$ and variance $v_1$. Similarly, $\mbox{Im(\emph{M})}$ is normally distributed with mean $m_2$ and variance $v_2$.
Thus, their joint distribution is given by
\be P(R,\Theta)=\frac{R}{2\pi \sqrt{v_1 v_2}}e^{-\frac{(R\cos\Theta-m_1)^2}{2v_1}-\frac{(R\sin\Theta-m_2)^2}{2v_2}},\label{joint}\ee where $R=|M|$ and $\tan\Theta=\mbox{Im(\emph{M})}/ \mbox{Re(\emph{M})}$. The required distribution function is obtained by an angular integration of eq.~(\ref{joint}). Performing the integration leads to the following infinite series of Bessel functions:

\be P(|M|)=\frac{|M|e^{-\alpha}}{\sqrt{v_1v_2}}\sum_{n=0}^{\infty}F_n(X,Y,Z)\cos[{2n\arctan(Z/Y})],\label {beck}\ee with

\be \alpha=\frac{m_1^2}{2v_1}+\frac{m_2^2+|M|^2}{2v_2}+\frac{v_2-v_1}{4v_1v_2}|M|^2\ee
and
\be F_n(X,Y,Z)=(-1)^n\epsilon_nI_n(X)I_{2n}(\sqrt{Y^2+Z^2}).\ee
 Here, $ X=(v_2-v_1)|M|^2/4v_1v_2$, $ Y=|M|m_1/v_1 $  and $Z=|M|m_2/v_2.$ $I_n(x)=i^{-n}J_n(ix)$ is the modified Bessel function of the first kind, $\epsilon_0=1$ and $\epsilon_n=2$ otherwise.
Figure \ref{fig3} shows the numeric spectrum of eq.~(\ref{U}) for $a=1.3$ (excluding the degenerate subradiant mode at $\Gamma=0$) as well as the corresponding calculated $P(\Gamma)$.
In the special case, considered earlier, where $a$ is an integer, it is easy to check that $m_1=m_2=0$ and $v_1=v_2=N/2$. Thus, eq.~(\ref{sc}) is recovered.

\section{V. Discussion}

The fundamental difference between the one and three-dimensional geometries, {\it i.e.,} the existence or the absence of a crossover between delocalized and localized photons, is due to the different nature of the coupling matrices. While $U$ falls off with the inter-atomic separation in the three-dimensional case, it is a periodic function of the inter-atomic distance in the one-dimensional geometry. Thus, the single atom limit is never reached.

Let us stress that eq.~(\ref{spec}) is valid for \emph{both} ordered and disordered media (in the case of an ordered system $M$ is not a random variable, but eq.~(\ref{spec}) still holds). Since the  disorder affects only two eigenvalues, namely $\lambda_\pm$,  $P(\Gamma)$ is comprised  of  $N-2$ vanishing eigenvalues regardless of  disorder, and $C=1$ for $N\gg 1$.  Therefore, cooperative effects and not  disorder is the mechanism that leads to photon localization in the case considered here. The same claim is valid in Dicke regime as well, since eq.~(\ref{eqc33}) holds for both ordered and disordered media. This unambiguous distinction between  the  contributions of disorder and cooperative effects to photon localization cannot be achieved in the three-dimensional geometry, where the role played by each of  these two mechanisms cannot be determined separately \cite{PRL2}.

The distribution of resonance widths in one-dimensional disordered media has also been studied in \cite{PRE}, where it has been shown that it follows a power law $P(\Gamma)\sim \Gamma^{-1}$ decay. The spectrum in  eq.~(\ref{spec}) does not, however,  obey this  power law. The difference stems from the fact that the authors of ref. \cite{PRE} have calculated $P(\Gamma)$ using the real part of the spectrum of the complex-valued Green matrix $\exp(ik_0 r_{ij})$, which describes propagation of a wave scattered by a dipole at $r_i$ to a dipole at $r_j$.  Here, we have taken a different approach \cite{PRL2, tallet1} and  studied the time evolution of the ground state population associated with the reduced atomic density operator of the system. As explained earlier, in our treatment one can interpret the eigenvalues of the real-valued matrix $\cos(k_0r_{ij})$ as the photon escape rates from the atomic gas. According to \cite{PRE}, the $P(\Gamma)\sim \Gamma^{-1}$  behavior can be interpreted as an unambiguous signature of Anderson localization of light in random systems. The fact that our result does not follow this power law supports the claim that cooperative effects and not  disorder is the mechanism that leads to photon localization in the case studied here.

It is interesting to compare these results to the two-dimensional case, where  $U_{ij}=J_0(x_{ij})$ \cite{book} where $x_{ij}$ is the dimensionless random distance between any two atoms. In this geometry,  when the atoms are close enough the Dicke limit is reached and eq.~(\ref{eqc33}) holds, as in the other geometries. In the opposite limit, $U$ can be approximated as $U_{ij}\simeq\sqrt{2/\pi x_{ij}}\cos(x_{ij}-\pi/4)$, and since it falls off with the square root of the inter-atomic separation, the single-atom limit can be reached. We  conclude that the absence of  single-atom limit is specific to the one-dimensional geometry.

Recently, the authors of ref. \cite{celardo} have studied the interplay of disorder and superradiance in a one-dimensional Anderson model in which all the sites are coupled to a common decay channel  with equal coupling strength. By diagonalizing the corresponding non-Hermitian Hamiltonian, the participation ratio of the eigenstates have been obtained. It has been shown that while subradiant states become localized as disorder increases, superradiant states remain delocalized. These results differ substantially from ours. The
difference stems from the fact that the authors in ref. \cite{celardo} have  considered an Anderson model with on-site disorder and assumed that the sites are coupled to the continuum with equal coupling
strength, while in our treatment  only the position-dependent continuum coupling is taken into account.

\section{VI. Summary}

We have studied cooperative effects in a one-dimensional random atomic system. By an analytic diagonalization of the Euclidean random matrix $U_{ij}=\cos x_{ij}$, where $x_{ij}$ is the dimensionless random distance between any two atoms, we  have calculated the photon escape rates from the gas for an arbitrary system size. We have shown that the single-atom limit is never reached and the photon is  always localized. This localization stems from long-range cooperative effects and not from disorder as expected on the basis of the theory of Anderson localization.

\section{Acknowledgements}
This work was supported by the Israel Science Foundation Grant No. 924/09.


\begin{thebibliography}{99}

\bibitem{dicke} R. H. Dicke, Phys. Rev. {\bf 93}, 99 (1954)

\bibitem{gross} M. Gross and S. Haroche, Phys. Rep. {\bf 93}, 301 (1982)

\bibitem{stephen} M. J. Stephen, J. Chem. Phys. {\bf 40}, 669 (1964)

\bibitem{lehmberg} R. H. Lehmberg, Phys. Rev. A {\bf 2}, 883 (1970)

\bibitem{scheibner} M. Scheibner, T. Schmidt, L.Worschech, A. Forchel, G. Bacher,
T. Passow, and D. Hommel, Nat. Phys. \textbf{3}, 106 (2007)

\bibitem{inouye} S. Inouye, A. P. Chikkatur, D. M. Stamper-Kurn, J. Stenger,
D. E. Pritchard, and W. Ketterle, Science \textbf{285}, 571 (1999)

\bibitem{baumann} K. Baumann, C. Guerlin, F. Brennecke, and T. Esslinger, Nature
\textbf{464}, 1301 (2010)

\bibitem{bienaime} T. Bienaime, N. Piovella, and R. Kaiser,
Phys. Rev. Lett. \textbf{108}, 123602 (2012)

\bibitem{kuraptsev} A. S. Kuraptsev, I. M. Sokolov, Ya. A. Fofanov,
Opt.  Spectrosc. \textbf{112}, 401 (2012)

\bibitem{wang} T. Wang, S. F. Yelin, R. Cote, E. E. Eyler, S. M. Farooqi, P. L.
Gould, M. Kostrun, D. Tong, and D. Vrinceanu, Phys. Rev. A
\textbf{75}, 033802 (2007)

\bibitem{PRL2} E. Akkermans, A. Gero and R. Kaiser, Phys. Rev. Lett. {\bf 101}, 103602 (2008)

\bibitem{tallet1} E. Ressayre and A. Tallet, Phys. Rev. Lett. {\bf 37}, 424 (1976)

\bibitem{milonni2} P. W. Milonni and P. L. Knight, Phys. Rev. A {\bf 10}, 1096 (1974)

\bibitem{vries} P. de Vries, D. V. van Coevorden and A. Langendijk, Rev. Mod. Phys. {\bf 70}, 447 (1988)

\bibitem{MPL} V. A. Marchenko and  L. A. Pastur, Mat. Sbornik USSR {\bf 1}, 457 (1967)

\bibitem{skipetrov}  E. Skipetrov and A. Goetschy, J. Phys. A:  Math. Theor. {\bf 44}, 065102 (2011)

\bibitem{beckmann} P. Beckmann, {\it Probability in Communication Engineering}, New York : Harcourt, Brace and World (1967)

\bibitem{PRE} F. A. Pinheiro , M. Rusek, A. Orlowski, and B. A. van Tiggelen, Phys. Rev. E {\bf 69}, 026605 (2004)

\bibitem{book} E. Akkermans and G. Montambaux, {\it Mesoscopic Physics of Electrons and Photons}, Cambridge University
Press (2007)

\bibitem{celardo} G. L. Celardo,  A. Biella, L.  Kaplan, and F. Borgonovi, Fortschr. Phys. {\bf 61}, 250 (2013)


\end{thebibliography}
\end{document}